\documentstyle[11pt,ihepconf,twoside,epsfig]{article}

\def\b{\bar}

\def\cA{{\cal A}}

\def\m{\mu}
\def\n{\nu}

\def\t{\tau}

\def\~{\tilde}

\def\bY3{\bar Y_{,3}}
\def\Y3{Y_{,3}}
\def\z{\zeta}
\def\Z{{\b\zeta}}
\def\Y{{\bar Y}}

\def\`{\dot}
\def\be{\begin{equation}}
\def\ee{\end{equation}}
\def\bea{\begin{eqnarray}}
\def\eea{\end{eqnarray}}

\def\fn{\footnote}

\def\cF{{\cal F}}

\def\mn{{\mu\nu}}

\begin{document}

\title{\bf \uppercase{Singular Strings \\ in the Rotating Astrophysical Sources:\\
a New Conjecture on the QPOs \\ [1mm] and Jet Phenomena}}

\author{{\bf Alexander Burinskii}\fn{
This talk is based on the collaboration with
 Emilio Elizalde, Sergi R. Hildebrandt and Giulio Magli.}\\
Gravity Research Group, NSI Russian
Academy of Sciences,
 115191 Moscow, Russia}

\date{}
\maketitle
\vspace*{5mm}
{\small
Stringy and disklike sources of the rotating compact astrophysical
objects are considered on the base of the Kerr geometry.
It is argued that analyticity of the Kerr solutions may result the
appearance of  singular strings, which may be the source of two
important astrophysical effects: the jets and QPOs phenomena.}
\vspace*{10mm}

\section{Introduction}

Most of the astrophysical works deal with the Kerr solution for
description of the rotating compact sources.
Description of such sources is usually based on the Kerr solution.
The Kerr geometry is a fundamental discovery which is related to
many physical systems, from the rotating black holes, neutron
stars and galactic nucleus in astrophysics till the structure of
spinning particles and fundamental solutions to the low energy
string theory.

Indeed, a tendency is now traced to an approaching of the physics
of the compact astrophysical objects and the physics of elementary
particles.
In particular, the  well known objects in the physics of
elementary particles,
such as the
bag models, superdense  matter, quarks and
color superconductivity, penetrate now in astrophysics,
demonstrating a Unity of physics.

In observational astrophysics this tendency is confirmed by the
problem of quasi-periodic oscillations, QPOs
phenomena (see for example \cite{McR,AKMR}).

Black hole binaries exhibit thermal and non-thermal components of
X-ray emission which vary widely in intensity.
The X-ray spectral and timing studies in the radio, optical and
gamma-ray diapasons display the stable
low frequency QPOs in the range 0.1-30 Hz and the
QPOs in the range 40-450 Hz.
Besides, there are observations of the
spectral lines which are in a relative stable harmonic relations.

There appears the problem of reconstruction of the models and
structure of
the compact objects basing on the spectral observations and timing.
The spectral analysis becomes one of the main tools of the modern
astrophysics resembling the atomic spectroscopy of the beginning
of the last century. However, so far there are very large
uncertainties for the corresponding physical models and we have
a risk here to conjecture a new approach to this problem.

Along with the QPOs problem,  there is also
the very old problem of the model of jet formations.

The aim of this work is to pay attention to some theoretical
effects which
are linked to analytic structure of the Kerr geometry and
reproduce some
features resembling the jet and QPOs phenomena.
At this stage we point merely out a qualitative
similarity, displaying a potential possible role of these effects,
so our conjectures are very far from the real estimations.

The presented model is based on the recent analysis of the
{\it aligned}
electromagnetic excitations on the Kerr background which was
performed
for investigation of the Kerr spinning particle \cite{BurAna}.
It was shown that the {\it aligned}
electromagnetic excitations of the Kerr geometry lead to the
appearance of two singular stringy structures:
\newpage

1) circular singular string

and

2) axial stringy system.

We conjecture that excitation of the circular string may be the
source of QPOs, while the appearance of the axial strings in some
cases may reproduce the jet phenomena. So,
 the QPOs and jet
problems may be connected to each other.

It may appear an obvious objection that
the electromagnetic excitations on the Kerr background were
investigated many times and by many authors.
The principal new feature of our treatment is the restriction by
the  electromagnetic fields which are {\it aligned} to the Kerr
principal null congruence. Physically, the aligned fields are
the only ones which
do not conflict with the twistorial analyticity of the Kerr
background, which means that they can be used for the formation
of the self-consistent solutions of the Einstein-Maxwell system
with a guarantee that the resulting metric (if exists) will be
of the same type as the Kerr solution, i.e. algebraically special
metric with a shear free and geodesic principal null
congruence.  This very natural demand turns out to be very
restricted indeed, and leads to unavoidable appearance of the
singular axial strings.

\section{Disklike sources of the Kerr solution}

In the papers \cite{behm,gurgur,BurBag} the smooth rotating
disk-like sources of the Kerr
and Kerr-Newman solutions were considered in the Kerr-Schild class of
metrics.

Note, that in the related previous works it was suggested by Sakharov,
Gliner and Markov  to replace the singularity of the non-rotating BHs by
the de Sitter source. This approach was developed in further
by Markov,Mukhanov\&Frolow, Israel\&Poison, Dymnikova, Magli et.al
(see references in~\cite{behm,BurBag}).

The Kerr-Schild metric has the form
\be g^{\mn}=\eta ^{\mn} - 2h k^\m k^\n ,
\label{ks}\ee
where $k^\m$ is the null
vector field $k_\m k^\m =0$ which is
tangent to the Kerr PNC (principal null congruence).

This is an extremely simple form of metric and one can wonder why this
form is able to describe the very complicated Kerr-Newman space-time.
To answer this question, first we mention  that the  function $h$ has the
form\fn{Here $M$ and $e$ are the mass and charge, and $a=J/M$ is a density of
angular momentum $J$ per mass unit.}
\be
h= \frac {Mr-e^2/2} {r^2 + a^2 \cos^2 \theta},
\label{h}
\ee
where the oblate coordinates $r,\theta$ are used on the flat
Minkowski background $\eta^{\mn}$, so the function $h$ is singular in
the focal points of this system $r=\cos \theta =0$, forming  the Kerr
singular ring.

\begin{figure}[ht]
\centering{\epsfig{figure=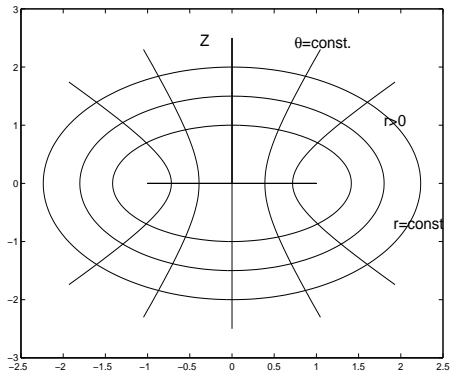,height=5cm,width=6cm}}
\caption{\small The oblate coordinate system $r,\theta$. Coordinate $r$ cover
the space twice: for $r>0$ and $r<0$. The focal points  are
$r=\cos \theta =0$.} \end{figure}

The Kerr singular ring is one of the most remarkable peculiarities of
the Kerr solution.  It is a branch line of space on two sheets: 
``negative''
($r<0$) and
``positive'' ($r<0$)
where the fields change their signs and directions.
The vector field $k^\m$
is the second remarkable structure of the Kerr geometry.
This field is lightlike or null, $k_\m k^\m =0$, and
forms a twisting principal null congruence (PNC).
It represents a vortex lightlike field propagating from the``negative'' sheet
of the Kerr space onto ``positive'' one. It is the very important structure.
Not only metric, but also vector potential
of the Kerr-Newman solution
\be A^\m = \frac {er}{r^2 + a^2 \cos^2 \theta} k^\m ,\ee
as well as the flow of radiation from the radiative Kerr source
$T_{rad}^{\mn} \sim \Phi (r, \theta) k^\m k^\n $
\cite{BurNst}
are determined by this vector field.
Its form,  shown on the fig.2,  shows that the most complicated part of
of the Kerr metric is concentrated in the form of the field $k_\m (x)$.

\begin{figure}[ht]
\centering{\epsfig{figure=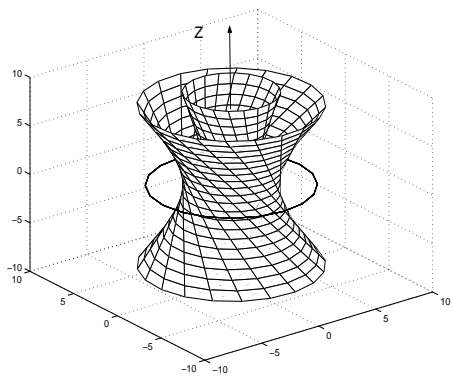,height=5cm,width=6cm}}
\caption{\small The Kerr singular ring and 3-D section of the Kerr principal null
congruence (PNC). Singular ring is a branch line of space, and PNC
propagates from ``negative'' sheet of the Kerr space to ``positive'' one,
covering the space-time twice. } \end{figure}

In the case $e^2 +a^2 >>m^2$, corresponding to parameters of
elementary particles, the horizons of the Kerr-Newman solution
disappear and the Kerr singular ring turns out to be naked.
This case is considered in the models of the Kerr
spinning particle \cite{behm,BurBag,BurAna,Isr,Bur0,IvBur,Lop,BurSup,BurStr}.
To avoid the problems with a twosheeted topology this
singularity may be covered by a (disklike) source~\cite{behm,BurBag,Lop}.
The naked Kerr singular ring is also considered in the models
of the Kerr spinning particle. In particular
as a waveguide providing a circular propagation of an electromagnetic or
fermionic wave excitations \cite{BurAna,Bur0,IvBur,BurTwo,BurSen}.
It was shown that the Kerr
singular ring represents a special type of the folded closed D-string
\cite{BurOri}.
 The value of charge $e$ is
small in astrophysical application, and one can consider as a critical case
of the BH formation the condition $|a/M|<1$.

In the approach of the papers \cite{behm,BurBag}, the core
of the Kerr source has the usual Kerr-Schild form of metric. However, the
function $h$ take the more general form
\be h=f(r)/ (r^2 + a^2 \cos^2 \theta ),\ee
where function $f(r)$ has the order $\sim r^4$  at $r=0$ to suppress
the singularity by $r=0$. Note, that for the nonrotating solutions this case
$f(r)=\alpha r^4 $ corresponds just to de Sitter core. The external metric is
chosen to be the vacuum Kerr solution.
Therefore, if a smooth function $f(r) $ interpolates between $ f_{core}
=\alpha r^4 $ and $f_{external}= Mr$ we obtain a smooth source of the Kerr
solution.

This matching may be conveniently displayed on the following
graphics.
\newpage

\begin{figure}[ht]
\centering{\epsfig{figure=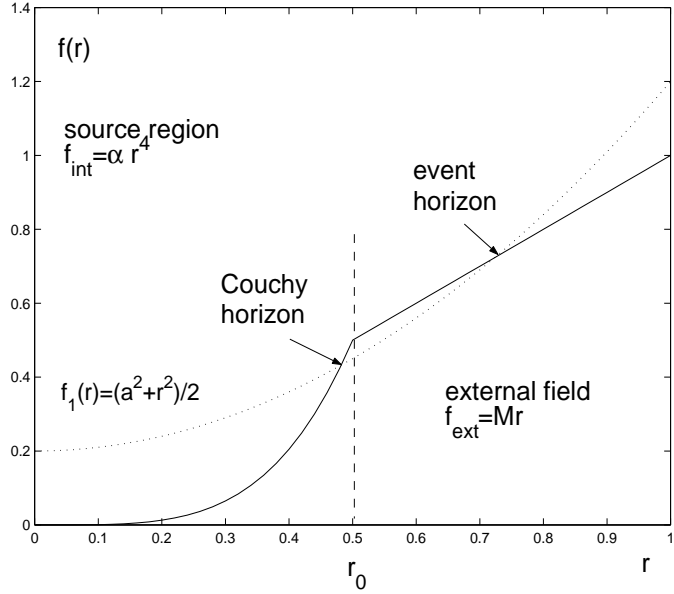,height=8cm,width=9cm}}
\caption{\small Matching the (rotating) internal ``de Sitter'' core with the
external Kerr-Schild field. The dotted line $f_1(r) =(r^2 +a^2)/2$
determines graphically the position of horizons as the roots of the equation
$f(r)=f_1(r)$.} \end{figure}

Note, that the condition of smoothness
does not depend from parameter $a$, so the change of metric due to rotation
is only controlled by the denominator of function  $h$ which is determined
by the relation of the oblate coordinate system to the Cartesian one

\be \cos \theta=z/r , \quad \sin \theta = \sqrt{\frac{x^2+y^2} 
{r^2 +a^2}} \, .
\ee

So the sources are represented in the form of the rotating disks with the
boundaries $r=r_0$.
Analysis of the structure of the electromagnetic field near the disk surface
has led to the conclusion that the matter of disk must have the
superconducting properties.
Electromagnetic field near the Kerr singular ring is singular. It means that
even if the total charge of the Kerr source is very small the electromagnetic
singularity is retained near the ring. It shows that the electromagnetic
effects can play very important role in the process of formation of the
Kerr disklike source.\\[-2mm]

{\bf Let us summarize the basic properties of the Kerr disk-like source:}
\vspace*{-1mm}
\begin{itemize}
\item the disk is oblate and rigidly rotating,
\item the rotation is relativistic,
\item  the stress-energy tensor has
 a special condensed vacuum state (de Sitter, flat or
anti de Sitter vacua).
\item  electromagnetic properties
of the disk are close to superconductor,
\item for the charged sources the strong magnetic and gravitational fields
are concentrated on the stringy board of the disk,
\item  the relation $J = M a$.
\end{itemize}

This model includes also the smooth analogs
of the known shell-like models.

\begin{figure}[ht]
\centering{\epsfig{figure=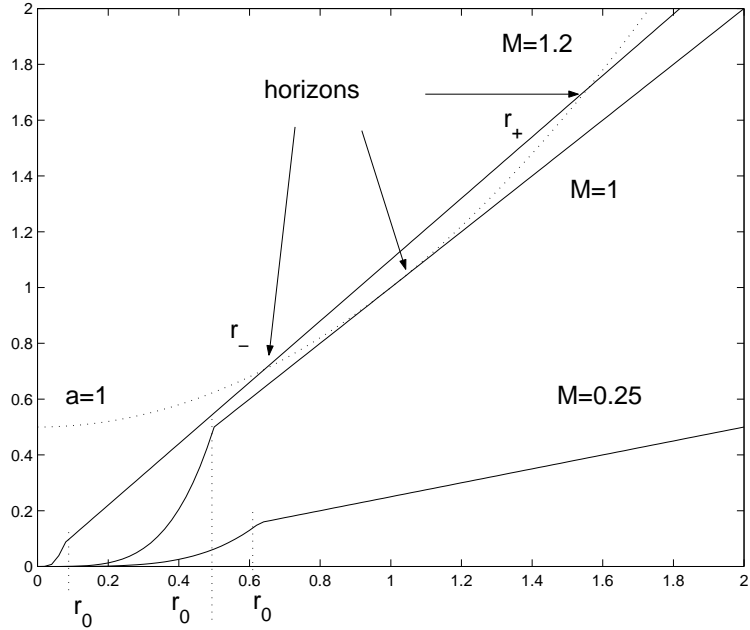,height=8.5cm,width=10cm}}
\caption{\small The sources with different masses $M$
and matter densities $\rho$.  Sources form the rotating disks with radius
$\sim a$ and thickness $\sim r_0$ which depends on the matter
density $r_0=(\frac{3M}{4\pi \rho})^{1/3}$. The formation of the black hole
horizons is shown for $a^2 <M^2$.} 
\end{figure}

In the limit of the infinitely thin disk a stringy singularity
is formed on the border of disk. This case corresponds to the considered by
Israel and Hamity the infinitely thin disklike source.

Twovaluedness of the metric and the field strengths was considered
in the field theory as an ``Alice'' property of the source
which was formulated at first for the cosmic ``Alice'' strings
\cite{Sch,Wit}.
The ``Alice'' phenomenon can be connected with the superconducting properties
of the sources where the ``negative'' sheet looks as a mirror image of the
``positive'' one.
This interpretation was also considered for the Kerr source \cite{BurStr}.
The ``Alice'' string is formed on the edge boundary of
the thin Kerr disk for $|a/M|>>1$.

\begin{figure}[ht]
\vspace*{-0,5cm}
\parbox{8 cm}{
\centering{\epsfig{figure=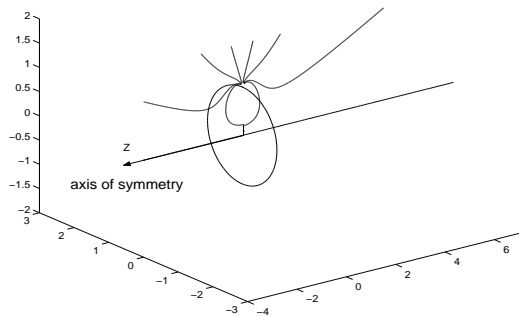,height=5cm,width=7cm, clip =}}}
\parbox{6cm}{\caption{\small Electric field strengths. } \par   }

 \end{figure}

\section{Stringy structures}

In the old paper \cite{Bur0}
the Kerr ring was considered as a gravitational
waveguide  caring the traveling electromagnetic waves which
generate the spin and mass of the Kerr spinning particle forming a
microgeon with spin.
It was conjectured \cite{IvBur} that the Kerr ring
represents a closed string, and the traveling
waves are the string excitations. It was noted in \cite{BurSen}
that in the axidilatonic version
of the Kerr solution the field around this ring is similar to the field
around a heterotic string, and recently, it was shown that the Kerr ring is a chiral D-string having
an orientifold world-sheet \cite{BurOri}.

\begin{figure}[ht]
\vspace*{-3mm}

\parbox{6 cm}{
\centering{\epsfig{figure=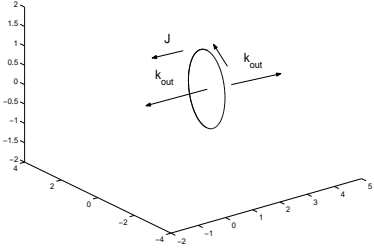,height=4cm,width=5cm, clip = }} 
}
\parbox{10 cm}{
\caption{\small Stringy
skeleton of the Kerr spinning particle. Circular string and axial stringy
system consisting of two semi-infinite strings of opposite chiralities.}
}
\end{figure}

\subsection{Axial stringy system}

Let us consider solutions for traveling waves -
electromagnetic excitations of the Kerr circular string.
The problem of electromagnetic excitations of the Kerr black hole
has been intensively studied
as a problem of the quasinormal modes. However, compatibility with the
holomorphic structure of the Kerr space-time put  an extra
demand on the solutions to be aligned to the Kerr PNC, which takes the form
$F^\mn k_\m=0$.
 The aligned wave solutions for electromagnetic fields
on the Kerr-Schild background were obtained in the Kerr-Schild
formalism \cite{DKS}. We describe here only the result referring
for details to the papers \cite{BurTwo,BurAxi}. Similar to the
stationary case \cite{DKS} the general aligned solution is
described by two self-dual tetrad components $\cF _{12} = AZ^2$
and $\cF _{31} = \gamma Z -(AZ),_1$, where function $A$ has the
form \be A= \psi(Y,\t)/P^2, \ee
 $P=2^{-1/2}(1+Y\Y)$, and $\psi$ is an arbitrary holomorphic function of
$\t$ which is a complex retarded-time parameter. Function
$Y(x) = e^{i\phi} \tan \frac {\theta} 2 $ is a projection
of sphere on a complex plane. It is singular
at $\theta=\pi$, and one sees that such a singularity will be present in
any holomorphic function $\psi (Y)$. Therefore, {\it all the aligned e.m.
wave solutions turn out to be singular at some angular direction $\theta$}.
 The simplest modes
\be
\psi _n = q Y^n \exp {i\omega _n \t}
\equiv q (\tan \frac \theta 2)^n \exp {i(n\phi + \omega _n \t)}
\label{psin}
\ee
can be numbered by index
$n=\pm 1, \pm 2, ...$,
which corresponds to the number of the
wave lengths along the Kerr ring.

Near the positive $z^+$ semi-axis we have $Y\to 0$  and
near the negative $z^-$ semi-axis  $Y\to \infty$.

Omitting the longitudinal components and the radiation field $\gamma$
one can obtain \cite{BurTwo,BurAxi} the form of the leading wave terms
\be \cF |_{wave} =f_R \ d \z \wedge d u  +
f_L \ d \Z \wedge d v ,
\label{cFLR}
\ee
where
$f_R = (AZ),_1 ; \qquad f_L =2Y \psi (Z/P)^2 + Y^2 (AZ),_1$
are the factors describing the ``left'' and ``right"  waves propagating
along the $z^-$ and $z^+$ semi-axis correspondingly.

The behavior of function $Z=P/(r+ia \cos \theta)$ determines a
singularity of the waves at the Kerr ring,
so the singular waves along the ring induce, via function $Y$,
singularities at the $z^\pm$ semi-axis. We are interested in the asymptotical
properties of these singularities.  Near the $z^+$ axis  $|Y|\to 0$,
and  by $r \to \infty $, we have $Y \simeq  e^{i\phi} \frac \rho {2r}$ where
$\rho$ is the distance from the $z^+$ axis. Similar, near the $z^-$ axis $Y
\simeq  e^{i\phi} \frac {2r} \rho  $ and $|Y|\to \infty$.
The parameter $\t=t  -r -ia \cos \theta$ takes near the z-axis the values $\t _+ = \t |_{z^+}=
t-z-ia,\quad \t _- = \t |_{z^-} =t+z +ia.$

The mode $n=0$ describes the stationary electromagnetic field of the
Kerr-Newman solution, so it does not contain a modulation of the
Kerr circular string and axial singularity is absent.

For $|n|>1$ the solutions contain the axial
singularities which do not fall of  asymptotically, but are increasing that
means instability.

The leading singular wave for $n=1$,
\be
\cF^-_1=\frac {4q e^{i2\phi+i\omega _{1} \t_- }} {\rho ^2} \ d \Z \wedge dv ,
\ee
propagates to $z=-\infty$ along the $z^-$ semi-axis.

The leading wave for $n=-1$,
\be
\cF^+_{-1}= -
\frac {4q e^{-i2\phi+i\omega _{-1} \t_+ }} {\rho ^2} \ d \z \wedge du ,
\ee
is singular at $z^+$ semi-axis and propagates to $z=+\infty$.
The described singular waves can also be
obtained from the potential
$\cA^\m= - \psi (Y,\t)(Z/P) k^\m $.
The $n=\pm 1$ partial solutions $\cA^{\pm} _n$ represent asymptotically
the singular plane-fronted e.m. waves propagating along
$z^+$ or $z^-$ semi-axis without damping.
The corresponding
self-consistent solution of the Einstein-Maxwell field equations
are described in \cite{BurTwo}. They are singular plane-fronted waves
 having the Kerr-Schild form of metric (\ref{ks}) with a constant vector
$k^\m$.  For example, the wave propagating along the $z^+$ axis has
$k^\m dx^\m= - 2^{1/2}du $).
The Maxwell equations take the form
$\Box \cA = J=0 , $ where $\Box$ is a flat D'Alembertian,
and can easily be integrated leading to the solutions
$ \cA ^+ = [ \Phi ^+(\z) + \Phi ^-(\Z) ] f^+(u) du, \qquad
 \cA ^- = [ \Phi ^+(\z) + \Phi ^-(\Z) ] f^-(v) dv,$
where $\Phi ^{\pm}$ are arbitrary analytic functions, and functions
$f^\pm $
describe the arbitrary retarded and advanced waves.
Therefore, the wave excitations of the Kerr ring lead to the appearance
of singular pp-waves which propagate outward along the $z^+$ and/or
$z^-$ semi-axis.

These axial singular strings are evidences
of the axial stringy currents, which are exhibited explicitly
when the singularities are regularized. Generalizing the field model to
the Witten field model for the cosmic superconducting strings \cite{Wit},
one can show \cite{BurAxi}  that these singularities are replaced by the
chiral superconducting strings, formed by a condensate of the Higgs field, so
the resulting currents on the strings are matched with the external gauge
field.

The case with two semi-infinite singular strings of opposite chiralities is
shown at the Fig.6.

In the cases $|n| > 1$ singularities cannot be stable since their
strength increases with distance. This case may describe
some type of jet, see Fig.7.
\vspace*{-4mm}
\begin{figure}[ht]
\parbox[b]{8 cm}{
\centerline{\epsfig{figure=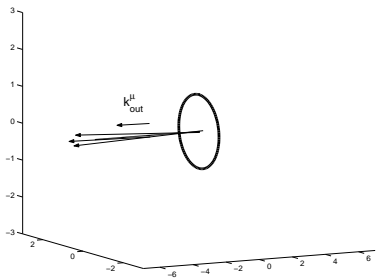,height=4cm,width=5cm}}
}
\parbox[b]{8 cm}{
 \caption{\small
 The Kerr circular singularity and a nonstable axial string corresponding to
$|n|>1$.}}
\vspace*{-0.2 cm}
\end{figure}

\section{Conclusion}
We have described here two new physical effects which are consequences of
the analytic structure of the Kerr geometry and follow from the analysis
of the exact solutions for
aligned electromagnetic excitations  on the Kerr background.  These solutions
have found application in the models of elementary particles \cite{BurAna},
and similar to the Kerr solution itself may find application in astrophysics
too.

First  of the effects is the possible
excitations of the stringy board of the disklike Kerr sources for $a>M$ with
formation of some resonances in a possible harmonic relation.

We should also note that the obtained aligned solutions are indeed
independent from the presence (or absence) of the horizon and, consequently,
the aligned electromagnetic oscillations do not forbidden for the black holes,
 too.

The second effect of the aligned solutions is the unavoidable appearance of
the axial singularities accompanied by outgoing traveling waves,
which may be source of the strong currents leading
to formation of astrophysical jets.

The most interesting is the fact that description of the above effects
is based on the natural assumption on the
analyticity of the Kerr background and does
not require implication of some additional assumptions.
This talk is based on the collaboration with E. Elizalde, G. Magli
and S.R. Hildebrandt. A.B. would also like to thank V. Frolov
and F. de Feliche for useful conversations.

\end{document}